\documentclass[pra,aps,letterpaper,longbibliography,twocolumn,12point]{revtex4-1}

\usepackage{graphicx}
\usepackage{xcolor}      
\usepackage{amsmath}

\newcommand{\bra}[1]{\langle #1|}
\newcommand{\ket}[1]{|#1\rangle}

\newcommand \be{\begin{equation}}
\newcommand \ee{\end{equation}}
\newcommand \bea{\begin{eqnarray}}
\newcommand \eea{\end{eqnarray}}

\newcommand{\adjoint}{^{\mbox{\footnotesize \dag}}} 

\begin{document}

\title{Speed, retention loss, and motional heating of atoms in an optical conveyor belt}
\author{G. T. Hickman and M. Saffman}
\affiliation{
Department of Physics, University of Wisconsin-Madison, 1150 University Avenue, Madison, Wisconsin 53706
}

\begin{abstract}
The problem of high-speed transport for cold atoms with minimal heating has received considerable attention in theory and experiment. Much theoretical work has focused on solutions of general problems, often assuming a harmonic trapping potential or a 1D geometry. However in the case of optical conveyor belts these assumptions are not always valid. Here we present experimental and numerical studies of the effects of various motional parameters on heating and retention of atoms transported in an optical conveyor. Our numerical model is specialized to the geometry of a moving optical lattice and uses dephasing in the density matrix formalism to account for effects of motion in the transverse plane. We verify the model by a comparison with experimental measurements, and use it to gain further insight into the relationship between the conveyor's performance and the various parameters of the system.
\end{abstract}

\date{\today}

\maketitle

\section{Introduction}
\label{sec:introduction}

Transport of cold neutral atoms or Bose Einstein condensates (BECs) with minimal heating is an important task in many atomic physics experiments. Common technologies used for this task include time-varying magnetic traps \cite{Nakagawa2005, Gupta2007, Purdy2010, Han2013, Pandey2019}, moving optical tweezers \cite{Beugnon2007, Couvert2008, Kim2016, Barredo2018, Samoylenko2020}, and moving optical lattices or optical conveyor belts \cite{Kuhr2001, Sauer2004, Brennecke2007}. The problem has also received attention from theorists, who have explored optimal transport protocols calculated by various means \cite{Torrontegui2013, Shumpei2009, Sorensen2016, Amri2019}. Analytical and numerical calculations become more cumbersome as the number of motional degrees of freedom is increased, however. Additionally, the wide variety of anharmonic potentials found in the various experiments cannot be accounted for exactly in a general way. As a result many theoretical investigations obtain rigorous or analytical results only by assuming that the confining potential is harmonic \cite{Murphy2009, Chen2011, Stefanatos2014, Ding2020}. Others simplify to a 1D problem by considering motion only along the direction of transport \cite{Torrontegui2011, Zhang2015}.

However, situations exist in which either one or both of these approximations do not hold. This is the case for trapped atoms when there is significant coupling between the transverse and longitudinal motions \cite{Torrontegui2012}. Also some experiments are performed in regimes in which trapping potentials are far from harmonic, either because of the intrinsic shape of the potential or because the typical energies of trapped atoms are comparable to the depth of the trap \cite{Ozeri1999, Poli2002, Balik2009}. In some of these cases new models will need to be developed that are specialized to the particular geometries. Here it will be important for theory work to take place side-by-side with experiment in order to validate the theoretical results.

Compared with other available transport technologies, optical conveyors are excellent tools for the precise positioning of atoms \cite{Kuhr2001, Schrader2001, Sauer2004, Nussmann2005, Dotsenko2005, Fortier2007, Dinardo2016}. They also have the unique ability to impart very large accelerations for high-speed transport \cite{Kuhr2001, Schrader2001}, and to quickly sort atoms into ordered arrays with well-defined spacings \cite{Miroshnychenko2006}. Many of the advantages of optical conveyors come from the high degree of longitudinal confinement provided by the lattice geometry. This can lead to very large differences between radial and longitudinal oscillation frequencies, which sets them apart from other transport system geometries. Atom lifetimes and noise sensitivities in optical lattices have been explored theoretically in \cite{Jauregui2001}, and more recently in \cite{Lu2020}, but heating rates in a moving lattice resulting from the transport procedure itself have not yet been carefully examined.

Particularly interesting would be an investigation of the effect of differing velocity waveforms on the retention and motional excitation of transported atoms. Motion in optical conveyors is often accomplished by driving the lattice velocity with a triangle-shaped waveform \cite{Schrader2001, Kuhr2003, Dotsenko2005, Langbecker2018}, even though this waveform is known to cause additional motional heating because of the discontinuities in the acceleration. Typically this source of heating is significant only when fast transport is required, using acceleration values that are close to the maximum allowed by the trap depth and the lattice geometry \cite{Schrader2001}. In some instances it suffices to simply transport the atoms more slowly, but in applications for which fast transport speeds are critical it would be beneficial to understand the performance improvements that can be obtained using a smooth velocity profile.

Here we present experimental and theoretical studies of heating and atom loss due to motion in an optical conveyor belt. In our experiments, atoms are loaded into an optical conveyor and transported back-and-forth across a fixed distance, and retention is measured using optical fluorescence imaging. On the theoretical side we introduce a 1D quantum numerical model for calculating the motional dynamics of an optical conveyor. The model is specialized to the case in which the frequencies of acousto-optic modulators (AOMs) are controlled digitally by performing discrete updates at a fixed rate, such as is commonly done with direct digital synthesizers (DDSs). Laboratory measurements of retention losses caused by parametric heating are used to calibrate the parameters of the model. We then compare the performances of triangle-shaped and smooth velocity waveforms by testing the retention as a function of transport speed, using both experimental measurements and numerical simulations. After verifying the accuracy of the model in this way, we use it to briefly investigate the relationship of motional heating and retention loss to the digitized update rates of the AOM frequencies. Finally, we use the model to study ground-state population dynamics in the moving trap. 

The structure of the remainder of this article is as follows: in Section \ref{sec:experiment} we review our experimental system and measurement protocol. In Section \ref{sec:simulation} we describe our simulation procedure, and simulation results are compared with experimental data in Section \ref{sec:theory_expt_comparison}. In Section \ref{sec:parametric_heating} we use the simulations to further analyze parametric heating and retention losses, and in Section \ref{sec:state_population} we use them to predict the effects of conveyor motion on atoms in the ground state of the lattice. We conclude in Section \ref{sec:conclusion}.

\section{Experiment}
\label{sec:experiment}

Our experimental system consists of a fiber Fabry-Perot cavity in an ultra-high vacuum (UHV) chamber with a background of $^{87}$Rb atoms at a pressure of roughly $5 \times 10^{-8}$ Torr. The high pressure is chosen to allow for a faster MOT loading rate and a shorter experiment cycle time. Atoms are cooled in a magneto-optical trap (MOT) at a distance of about $2~\rm  mm$ from the cavity and transferred to a far-detuned optical lattice. The MOT has a typical diameter of close to $80~\mu$m, overlapping with about 160 local minima of the lattice. The number of atoms loaded per cycle is on the order of $1000$, so the average number of atoms trapped in each lattice site is about $10$. This leads to a typical atomic density in the lattice on the order of $10^{12}~{\rm cm}^{-3}$.

Light for the lattice is produced by a 1064 nm single-frequency fiber laser (IPG Photonics model \# YLR-20-1064-LP-SF). The laser output is split by a polarizing beamsplitter, after which each of the two beams passes through an AOM. The AOMs are driven by two DDSs sharing a common clock (Sinara 4410 ``Urukul''). The lattice beams are focused onto the center of the fiber cavity with waists of $w_x \approx 30~ \mu$m and $w_y \approx 18~ \mu$m in the horizontal and vertical directions, respectively, and with Rayleigh lengths of $z_{\text{R},x}\approx2.7~\rm mm$ and $z_{\text{R},y}\approx1.0~\rm mm$. The smaller value of $w_y$ allows the trap beams to fit more easily between the cavity fibers. At the position of the MOT the beam waist sizes in the two transverse directions are both approximately equal to $40~\mu$m. The trap depth at that position is about $340~\mu$K.

Before beginning an experiment the temperature of atoms in the dipole trap is measured using the time-of-flight technique. For the experiments shown here the temperature was found to be $T=40~\mu$K. Due to the geometry of our setup the temperature is only measured in the directions orthogonal to the conveyor axis. The initial longitudinal motional temperature before transport is taken to be equal to the transverse motional temperature measured in this way.

\begin{figure}[t]
\includegraphics[width=3.4in]{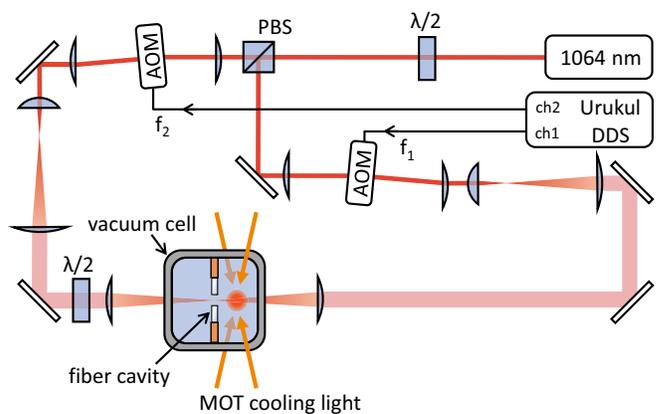}
\caption{(Color online) Optics layout for the optical conveyor experiment. (AOM) acousto-optic modulator, (DDS) direct digital synthesizer.}
\label{fig:opt_layout}
\end{figure}

After atoms are loaded into the lattice, frequency ramps are applied to the AOMs in order to produce a frequency difference $\Delta f \equiv f_\text{1}-f_\text{2}$ between the counterpropagating lattice beams, where $f_\text{1}$ and $f_\text{2}$ are the AOM driving frequencies. Starting from a common value of $f_\text{1}=f_\text{2}=80~$MHz, one of the frequencies is shifted up while the other is shifted down. The intensity maxima of the lattice then move with velocity $v=\lambda\Delta f/2$, enabling the lattice to carry the atoms into the field mode of the fiber cavity \cite{Kuhr2001, Schrader2001}. This setup has been designed for future experiments in quantum networking \cite{Wade2016}, though for the present work we do not make use of the cavity. The AOMs are used in a single-pass configuration to minimize optical losses. Their crystal facets are imaged onto each other by the focusing optics, and the modulators are oriented such that the two counterpropagating beams remain overlapped when a frequency ramp is applied. The setup of the conveyor optics is illustrated in Figure \ref{fig:opt_layout}.

DDS frequency ramps are implemented in discrete steps, the size and frequency of which can strongly affect the performance of the optical conveyor. In order to reach a wider range of stepping frequencies, we operate our DDS's in random access memory (RAM) mode. In this mode a list of output frequencies, amplitudes, or phases are stored in memory local to the DDS chip and can then be played back at rates of up to 250 MHz. MOT cooling light frequencies and intensities, magnetic fields, experiment timings, and DDS amplitudes and frequency ramps are controlled using hardware and software made available through the ARTIQ project \cite{Artiq}.

\begin{figure} [t]
\includegraphics[width=3in]{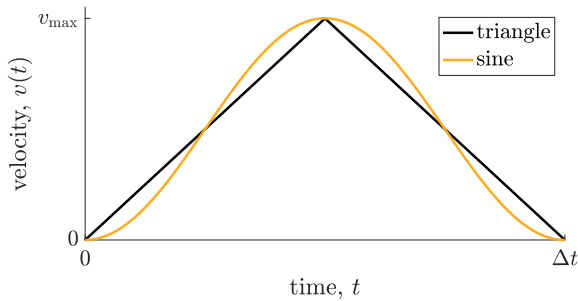}
\caption{Triangle and sine velocity profiles for the optical conveyor. The parameter $\Delta t$ indicates the total one-way travel time, and $v_\text{max}$ is the maximum velocity during transport.}
\label{fig:profiles}
\end{figure}

At the beginning of each measurement $^{87}$Rb atoms are collected in a MOT for $350$ ms. They are then loaded into the optical lattice, while the detuning of the cooling light is gradually increased over the course of $2.5~$ms. The cooling light is then switched off and the relevant motional experiment is performed. Digital control of the AOM driving frequencies allows us to apply almost any arbitrarily-shaped velocity profile to the conveyor. We perform experiments using both triangle- and sine-shaped profiles, illustrated in Fig.~\ref{fig:profiles}. After the experiment, retention is measured by turning on the MOT cooling light and imaging the fluorescence using an EMCCD camera (Andor Luca).

Whereas for this work we are concerned primarily with the effects of the transport process, in fact the retention in the optical lattice depends also on various other sources of heating, including technical noise and beam pointing instabilities \cite{Jauregui2001, Lu2020}. In order to remove these effects from the data each experiment is repeated twice: once with the lattice held stationary for the entire measurement, and a second time with the conveyor moving according to the predetermined pattern. The retention for the second experiment is then normalized against that of the first. Data acquired using this procedure is presented in Section \ref{sec:theory_expt_comparison}.

\section{Simulation}
\label{sec:simulation}

Here we review the details of our quantum simulation for modeling retention and heating in an optical conveyor. Our goal is to develop a 1D model that correctly accounts for effects of the 3D atomic motion. We begin with the full 3D potential of our optical lattice
\begin{equation}\begin{aligned}
U_{\rm full}(x,y,z) = &-U_{\rm f,0}(z) \cos^2 (k z) ... \\
    ... &e^{-2\left(x^2/w_x^2(z) + y^2/w_y^2(z) \right)} ,
\label{eq:lattice_potential}
\end{aligned}\end{equation}
where $k=2\pi/\lambda$ is the wavenumber of the dipole trap laser with $\lambda=1064$ nm, $z$ is the longitudinal coordinate, and $x$ and $y$ are the transverse coordinates. The beam waists in the two transverse directions $w_x(z)$ and $w_y(z)$, and the depth of the lattice $U_{\rm f,0}(z)$ are in general functions of the longitudinal position $z$. However we assume here for simplicity that transport distances are much smaller than the Rayleigh length of the trapping fields, so that these can all be approximated by constants. For larger distances the variations as a function of $z$ may need to be taken into account.

Next we reduce the problem down to a single dimension. However, it is not sufficient simply to neglect the transverse degrees of freedom. Because of transverse motion the time-averaged values of $x^2$ and $y^2$ are generally greater than zero. This leads to a reduction in the effective depth of the potential as seen by the longitudinal motion of the atoms, due to the presence of the exponential in Eq. (\ref{eq:lattice_potential}). The effective potential depth that is obtained after taking this reduction into account we call $U_0$. We treat $U_0$ as a free parameter of the model. Later in Section \ref{sec:theory_expt_comparison} we discuss how its value can be determined. 

As an additional simplification we neglect all of the potential wells of the periodic lattice except for one, in order to make the model more amenable to numerical simulations. We are left with the 1D potential
\begin{equation}
U(z) =-U_\text{0} \Pi (z) \cos^2 ( k z ) ,
\label{eq:single_site_potential}
\end{equation}
with $\Pi ( z ) \equiv H_s ( z + \pi/(2 k) ) - H_s ( z - \pi/(2 k) ) $, where $H_s$ denotes the Heaviside step function. Close to the ground state this potential can be approximated by that of a harmonic oscillator with angular frequency $\omega_0=k \sqrt{2 U_\text{0}/m}$. Note that the true lattice potential admits of an energy level band structure, whereas the potential of Eq. (\ref{eq:single_site_potential}) supports only a finite number of discrete eigenstates. Simplifying the lattice to a single site carries with it the risk of incurring errors due to the neglect of the complex level structure within the bands \cite{Jauregui2001}. Effects of the finite bandwidths can nevertheless be reintroduced phenomenologically by including a line broadening mechanism such as dephasing, as will be done later in this section.

The time-independent Schr\"{o}dinger equation is then solved to find the eigenstates and eigenenergies of an atom trapped in the potential $U\left(z\right)$. From the finite number $N$ of bound energy levels, the lowest $N_{\rm eff}$ are chosen to be included in the calculation. $N_{\rm eff}$ is treated as a free parameter when fitting to experimental data. An initial density matrix $\rho_{\rm init}$ is constructed in the basis of these eigenstates with
\begin{equation}
\rho_{\rm init} = \sum\limits_{i=1}^{N_{\rm eff}} p_i \ket{\psi_i}\bra{\psi_i},
\label{eq:rho_init}
\end{equation}
where $\ket{\psi_i}$ is the $i$'th eigenstate and the $p_i$ are the eigenstate occupation probabilities determined by the initial thermal state.

The time evolution of the system is calculated using propagator matrices. The discrete velocity boosts resulting from the digital implementation of the AOM frequency ramps are simulated by repeatedly applying an operator $T_\text{b}\left(\delta v \right)$ in a series of steps, where the operator is a function of the size of the velocity boost $\delta v$ given to the trap during that step. These operators are not trace-preserving, which accounts for the fact that each boost has some chance to eject the atom from the trap.

The matrix elements of $T_\text{b} (\delta v )$ are calculated with the inner product
\begin{equation}
T_{\text{b}(i,j)}(\delta v )= \int \psi_i^* ( z ) \psi_j ( z )  e^{- \imath m  \delta v \cdot z / \hbar} dz,
\label{eq:Uboost_matrix_elements}
\end{equation}
where $m$ is the mass of a $^{87}$Rb atom. The minus sign in the exponent accounts for the fact that the reference frame receives the boost by $\delta v$, rather than the atom. In general $\delta v$ is not constant over the course of an experiment, but calculating $T_\text{b}$ for every relevant value of $\delta v$ in this case would be too computationally intensive. Instead, we evaluate Eq. (\ref{eq:Uboost_matrix_elements}) for a reasonably large number of values of $\delta v$, usually 60 of them, and interpolate between the results. The differences between operator matrix elements found in this way and those obtained by direct calculation are typically of order $10^{-6}$ or less, which is sufficiently precise for our purposes. For the time between velocity boosts the dynamics are governed by the Hamiltonian for a single atom in a stationary lattice site,
\begin{equation}
H=\sum\limits_{i=1}^{N_{\rm eff}} E_i \ket{\psi_i}\bra{\psi_i},
\label{eq:Hamiltonian}
\end{equation}
where $E_i$ is the energy of eigenstate $i$. The motion during these periods is calculated by applying the propagator for free evolution,
\begin{equation}
T_\text{f}(\delta t)=e^{\imath H \delta t/\hbar},
\end{equation}
with $\delta t=1/f_{\rm DDS}$ the time step between velocity boosts due to DDS frequency updates, and $f_{\rm DDS}$ the DDS update rate. At this point it is straightforward to write the density matrix $\rho_i$ after step $i$ in terms of the density matrix $\rho_{i-1}$ at the end of the previous step,
\begin{equation}
\rho_i=T_\text{f}(\delta t) {T_\text{b}}(\delta v_i) \rho_{i-1} {T_\text{b}}\adjoint(\delta v_i)  {T_\text{f}}\adjoint(\delta t).
\label{eq:one_step_part1}
\end{equation}
Here $\delta v_i$ is the velocity boost given during step $i$. This relationship can be applied recursively to determine the density matrix evolution over the course of the full motion. However, the model so far accounts only for the effects of 1D motion in the longitudinal direction.

In order to accurately describe the motion of atoms in a real 3D potential the treatment must be expanded. In general, motion in a 3D anharmonic trap involves some degree of mixing between the motional states for the 3 dimensions. For a typical 1D optical lattice this effect is partly suppressed by the large difference between the longitudinal and transverse trap frequencies. In this case motion in the transverse directions has the effect of inducing a slow variation onto $\omega_0$ and onto the effective trap depth $U_\text{0}$, which acts as a kind of line broadening. We model this by applying dephasing when calculating the evolution of the 1D motional density matrix.

Dephasing can also be used to account partly for the effects of the finite bandwidths that appear in the energy level structure of a 1D lattice, as was discussed previously, though in the case of our experiment we suspect that this effect is smaller than that of the transverse motion. Collisions between trapped atoms can be accounted for as well, however given the densities used in our experiment the typical rate for these collisions is on the order of a few Hz \cite{Arpornthip2012}. This is negligible in comparison with the transverse motional effects.

Applying the dephasing in the simplest possible way, by using an identical dephasing rate for all states, improves greatly the agreement between the theoretical calculation and experimental data. However the best fits have been obtained using a heuristic model in which the dephasing rate is taken to be larger for states with larger departure from harmonicity. The modeled dephasing rate $\gamma_i$ for state $\ket{\psi_i}$ with energy $E_i$ is given by
\begin{equation}
\gamma_i = \begin{cases}
	 2\gamma_0 \frac{\left( \hbar \omega_0 - E_i + E_{i-1} \right)}{\textrm{Max}_j\left[ \hbar \omega_0 - E_j + E_{j-1} \right]} & i>1  \\
	 2\gamma_0 \frac{\left( \hbar \omega_0 - 2E_1 \right)}{\textrm{Max}_j\left[ \hbar \omega_0 - E_j + E_{j-1} \right]} & i=1
	 \end{cases},
\label{eq:deph_rate}
\end{equation}
where $\textrm{Max}_j[...]$ represents the maximum over $j$ and $\gamma_0$ is a free parameter representing roughly the average dephasing rate over all states. A dephasing matrix $M_\text{d}$ is constructed with elements
\begin{equation}
M_{\text{d}(i,j)} = \begin{cases}
	e^{-(\gamma_i + \gamma_j) \delta t / 2} & i \neq j \\
	1 & i=j
	\end{cases}.
\label{eq:Md}
\end{equation}
We apply the dephasing by performing element-by-element multiplication of $M_\text{d}$ with the density matrix $\rho$ after each time step $\delta t$. This accounts for the exponential decay of coherence that would be obtained from a typical master equation solution with dephasing terms. With the inclusion of this effect the relation for the evolution becomes
\begin{equation}
\rho_i=T_\text{f}(\delta t) {T_\text{b}}(\delta v_i) \rho_{i-1} {T_\text{b}}\adjoint(\delta v_i)  {T_\text{f}}\adjoint(\delta t) \times M_\text{d},
\label{eq:one_step_part2}
\end{equation}
where the `$\times$' symbol indicates element-by-element multiplication. As with the trap depth $U_0$, we assume that $\gamma_0$ remains constant over the course of the motion. This may not be valid when atoms are transported over distances much larger than the Rayleigh length, in which case it might be necessary to incorporate a time-dependent value for $\gamma_0$.

In most cases the method described above is sufficient to predict the retention of an atom in our optical conveyor. However, some modification is required in the case of very large accelerations for which the potential Eq. (\ref{eq:single_site_potential}) is strongly perturbed. The density matrix that we use is constructed from the eigenstates of a stationary lattice site, and for a strongly accelerated lattice these are no longer the correct eigenstates. The dephasing must be applied in the eigenbasis of the accelerated potential. Otherwise it has the undesirable effect of mixing together the populations of the accelerated eigenstates, adding a nonphysical source of heating.

To determine the correct basis we begin by writing the effective potential for a single accelerating lattice site:
\begin{equation}
U_\text{a}(z) =-U_\text{0} \Pi(z) \cos^2 ( k z ) + maz,
\label{eq:accelerating_potential}
\end{equation}
where $a$ is the acceleration of the lattice. One effect of the addition of the acceleration term in Eq. (\ref{eq:accelerating_potential}) is a shift of the position of the trap minimum by an amount
\begin{equation}
\Delta z (a) = -\frac{1}{2k}\arcsin \left( \frac{ma}{k U_\text{0}} \right).
\end{equation}
Here $\Delta z$ is the final position of the potential minimum, since the stationary lattice is defined so that its minimum occurs at $z=0$. To approximate the accelerated basis states we use the eigenstates of the original stationary potential after displacement by $\Delta z$, $\psi_i' (a, z) \equiv \psi_i (z - \Delta z(a))$ for $i \in \{1,...,N_{\rm eff}\}$, with eigenfunctions of the accelerated Hamiltonian designated by a prime. We then calculate an operator $T_\text{x}(a)$ to transform between bases. Its matrix elements are determined by
\begin{equation}
T_{\text{x}(i,j)}(a) = \int \psi_i'^* (a, z ) \psi_j ( z ) dz.
\label{eq:Ux}
\end{equation}
The acceleration $a$ is proportional to $\delta v$, so in general it is not constant over the course of an experiment. In order to make efficient use of computational resources we calculate $T_\text{x}(a)$ for a modestly large number of values of $a$ and interpolate between the results, similar to what was done in the case of $T_\text{b}(\delta v )$.

The simulation begins by assuming an initial thermal distribution at a temperature of $40 ~\mu$K and proceeds to step through each of the discrete velocity boosts for the given experiment. During each step the operations performed are as follows: First, the lattice velocity boost transformation $T_\text{b}(\delta v)$ is applied, followed by the free evolution propagator for a time of length $\delta t$, $T_\text{f}(\delta t)$. After this the basis is transformed into the frame of the accelerating lattice. Then the dephasing matrix is applied using element-by-element multiplication, and finally the frame is transformed back to the basis of the stationary lattice. Including all terms, the relation for the evolution of the density matrix can be written as
\begin{equation}\begin{aligned}
\rho_i= &~{T_\text{x}}\adjoint (a_i) \Big[ \Big[ T_\text{x}(a_i) T_\text{f}(\delta t) {T_\text{b}}(\delta v_i) \rho_{i-1} ... \\ 
	... &  {T_\text{b}}\adjoint(\delta v_i)  {T_\text{f}}\adjoint(\delta t) {T_\text{x}}\adjoint (a_i) \Big] \times M_\text{d} \Big] T_\text{x}(a_i).
\label{eq:one_step}
\end{aligned}\end{equation}
Here $a_i = \delta v_i / \delta t$ is the effective acceleration during step $i$. At the end of the simulation the diagonal elements of the final density matrix are stored for analysis.

\section{Theory and Experiment Comparison}
\label{sec:theory_expt_comparison}

For our experiments, an ensemble of atoms is loaded into the optical conveyor and transported back-and-forth 10 times across a distance of $\Delta x = 0.2$ mm. The short distance is used in order to allow testing of a larger range of DDS update rates $f_{\rm DDS}$, given the limited RAM storage space of our DDSs, and also to ensure that $\Delta x \ll z_{{\rm R},x/y}$. The multiple trips serve to increase the losses, particularly for the triangle wave velocity profile which introduces additional heating with each change of direction. This helps to make the loss features easier to measure and to compare with theory.

To begin the experiment, a set of measurements are performed to determine the retention for fixed value of the one-way trip time $\Delta t$, but variable $f_{\rm DDS}$. After this another set of measurements is made with variable $\Delta t$ but a fixed number of DDS update steps $N_\text{steps}$. The fixed number of steps is used in this case, rather than a fixed update frequency, for simplicity because of the details of the experimental implementation. The relationship between retention and $\Delta t$ is measured twice, once with a triangle-shaped velocity profile and again with a sine-shaped profile. Each experiment is repeated 125 times and results are averaged to improve the signal-to-noise ratio.

A 1D lattice dipole trap such as ours can sustain a maximum acceleration of $a_{\rm max}=U_0 k/m$ before the minima of the trapping potential vanish \cite{Schrader2001}. For our experiment $a_{\rm max}=1.43\times 10^5$ m/s, using the effective trap depth $U_\text{0}/k_{\rm B}=254~\mu$K obtained from the simulation parameters, with $k_{\rm B}$ the Boltzmann constant. For a sine-shaped velocity waveform the acceleration $a_{\rm max}$ is reached at a transport time of $\Delta t_\text{min}=0.094~$ms. When varying $\Delta t$, values close to this number were chosen in order to probe the conveyor's performance near the absolute limit of its speed.

After the measurements, our numerical model is tested against the experiment. The model has three free parameters: the effective trap depth $U_0$, the number of included bound states $N_{\rm eff}$, and the dephasing parameter $\gamma_0$. In the following we will briefly consider each of these parameters and offer some guidelines for optimizing their values.

A typical procedure for measuring the effective depth $U_0$ of an optical trap involves the introduction of some parametric excitation and subsequent measurement of trap retention for a variety of excitation frequencies. In our case this is done by varying the DDS update frequency $f_{\rm DDS}$. Typically the trap frequency $f_{\rm trap}=\omega_0/(2\pi)$ is inferred by identifying the fundamental and/or harmonic resonances by inspection, and the trap depth is calculated from $f_{\rm trap}$. However this procedure produces only an approximate result, as it neglects the anharmonicity of the trap. In particular, for an optical lattice the losses tend to occur for excitation frequencies $f_{\rm DDS}$ somewhat smaller than $f_{\rm trap}$. This results from the fact that the energy spacing between adjacent levels decreases with increasing energy. When $f_{\rm DDS} = f_{\rm trap}$ the parametric excitation mixes together the populations of the lower-lying states, but it has little effect on the more energetic states because of the larger detuning. It is only when $f_{\rm DDS}<f_{\rm trap}$ that significant losses are observed, since atoms in higher-lying states are more easily excited out of the lattice. This fact was noted as well in \cite{Jauregui2001}.

A robust method for determining the value of $U_0$ is to infer it approximately by inspection of measured parametric excitation data, and then to refine the number more carefully by comparing the data with simulation results. It should be kept in mind though that the effective trap depth $U_0$ obtained by fitting our model to measured data will in general be smaller than the full depth $U_{\rm f,0}$ of the actual optical trap. This results from the additional energy contained on average in the transverse motional degrees of freedom, as was mentioned in Section \ref{sec:simulation}. The size of the discrepancy will vary from one experiment to another due to the differing trap geometries and levels of transverse excitation. For our system, a comparison with 3D Monte Carlo simulations indicates that $U_{\rm f,0}$ is about $35\%$ larger than our fitted value for $U_0$.

\begin{figure}[t]
\includegraphics[width=3.4in]{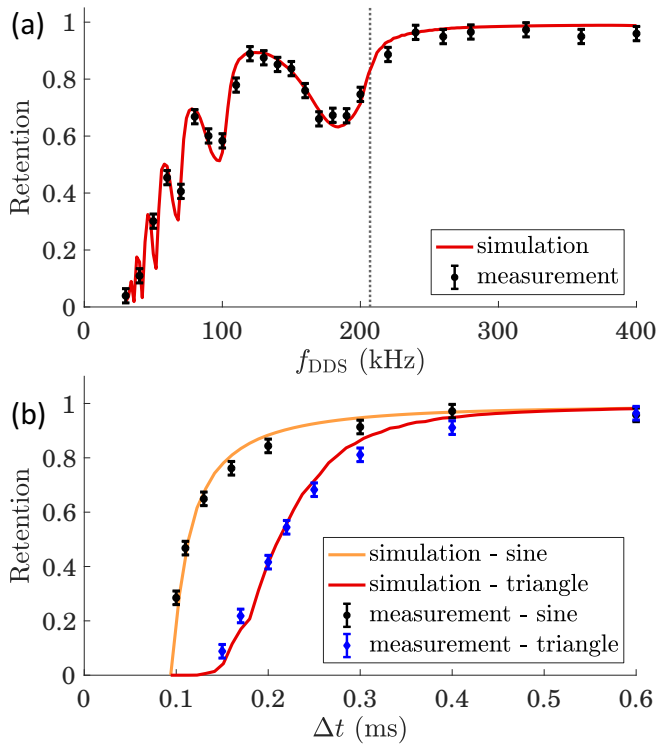}
\caption{Experiment and simulation results. The two agree very well. (a) Retention vs. DDS update frequency $f_{\rm DDS}$, using a sine-shaped velocity waveform. The one-way trip time is held constant at $\Delta t=1$~ms. The vertical dashed line marks the location of the fitted trap frequency $f_{\rm trap}=207$ kHz. (b) Retention vs. $\Delta t$, with the total number of DDS frequency steps held constant at $N_{\rm steps}=200$. Note that the simulations shown in (b) are performed without any adjustment of the free parameter values found from (a). The parameter values used here are: $U_0/k_{\rm B}=254~\mu$K, $N_{\rm eff}=28$, and $\gamma_0/2\pi=1.67$ kHz.}
\label{fig:results_main}
\end{figure}

When optimizing $N_{\rm eff}$, for our experiment the best agreement is typically obtained using values close to $N_{\rm eff} \approx 0.85 N$. The need to use $N_{\rm eff} < N$ arises from the nonzero transverse motion of atoms. As was mentioned in Section \ref{sec:simulation}, motion in the transverse directions produces slow variations of the effective trap frequency $\omega_0$ and depth $U_0$. This leads to an effective dephasing, as was discussed previously, but also to a reduction in the effective number of bound states. Higher-lying longitudinal bound states which exist when the atom is close to the optical axis can disappear as it moves farther away. Atoms excited into these states quickly escape from the trap, so the simulation more accurately reproduces experimental data by neglecting them. Quantum tunnelling may also play a role in the depletion of the populations of the more highly excited states \cite{Jauregui2001}. Because of the relatively short timescales of our experiments though, and because a large number of successive tunneling events would be required to remove an atom from the lattice, we suspect that this contribution is smaller in our case.

The dephasing parameter $\gamma_0$ can be determined empirically, but it is helpful to first make an order-of-magnitude estimate. To do this we start by looking at the phase relationships between energy eigenstates, with some simplifying assumptions. Coherence between the lattice eigenstates is maintained when the relative phases between adjacent states obey the relationship,
\begin{equation}
\Delta \phi_{(i,i-1)} \equiv \phi_i(t)-\phi_{i-1}(t) = \omega_0 t/ \hbar.
\label{eq:Delta_phi}
\end{equation}
We have assumed a harmonic potential here for simplicity. Due to transverse motion the actual phases gradually wander away from these values. To approximate the dephasing time $\tau \equiv 1/ \gamma_0$, we estimate the amount of time required for the actual relative phase and the ideal one of Eq. (\ref{eq:Delta_phi}) to accumulate a difference of about $\pi$, given our experimental conditions. Instead of treating the full 2D transverse trajectory for this, we consider only a single dimension of motion. This motion will introduce oscillations onto $\omega_0$, which then becomes time-dependent. Suppose that these oscillations are approximately sinusoidal, so that $\omega(t) = \omega_0+ \Delta\omega(t)$, with
\begin{equation}
\Delta\omega(t) \approx \Delta\omega_0 \sin(2\omega_r t),
\label{eq:Delta_omega_of_t}
\end{equation}
where $\omega_r \approx 2\pi \times 1.6$ kHz is the angular frequency for transverse motion in our lattice. The amplitude $\Delta\omega_0$ now needs to be determined.

The transverse motional energy we take to be equal to $k_BT$, with $T=40~\mu$K the measured temperature of the atoms. The transverse motion causes the effective longitudinal trap depth to oscillate about its average value. The average we take to be equal to $U_0$, which is only approximately correct but is good enough for an initial estimate. Then the minimum and maximum values for the trap depth are $U_{\rm 0,min} \approx U_0-k_{\rm B}T/2$ and $U_{\rm 0,max} \approx U_0+k_{\rm B}T/2$. Subsequently the trap frequency $\omega(t)$ oscillates between $\omega_{\rm max}=k \sqrt{2 U_\text{\rm 0,max}/m}$ and $\omega_{\rm min}=k \sqrt{2 U_\text{\rm 0,min}/m}$, so we can use $\Delta\omega_0=\omega_{\rm max} - \omega_{\rm min}$. Finally we approximate $\tau$ by integrating Eq. (\ref{eq:Delta_omega_of_t}),
\begin{equation}
\pi \approx \int_0^\tau \lvert \Delta\omega(t) \rvert dt,
\label{eq:tau}
\end{equation}
and then take $\gamma_0 \approx 1/\tau$. The absolute value is used to ensure that a solution always exists. For our experiment we obtain $\gamma_0/2\pi \approx 1.6$ kHz, which is very close to the final fitted value of $1.67$ kHz. Given the number of approximations used we suspect that the very good agreement between these two numbers is mostly accidental. In general we expect them to agree roughly to within a factor of two.

After approximate values have been obtained for each of the simulation parameters, fine adjustments are made by comparing simulation results with the measured retention vs. $f_{\rm DDS}$ data. Once the best parameter values have been found, a second set of simulations are run using these predetermined values. In this second set we calculate retention as a function of the one-way travel time $\Delta t$, for both sine-shaped and triangle-shaped velocity waveforms.

Our experimental measurements and fitted simulation results are shown in Fig.~\ref{fig:results_main}. The simulation obtains quite good quantitative agreement with experiment. This success it owes largely to the parameter $\gamma_0$, which compensates for transverse motional effects. By contrast, simulations with $\gamma_0=0$ produce very poor agreement - for example, no clear resonances are observable in plots of retention vs. $f_{\rm DDS}$.

The measurement and simulation results agree that the retention of the sine velocity profile noticeably outperforms the triangle profile. In our case, for a fixed retention of 90\% the sine profile allows transport to be performed about 25\% faster than the triangle. Simulations tested using a range of other smooth velocity profiles showed at best very similar performance to that of the sine. So, though the sine shape may not be quite optimal, it seems that it will be sufficiently close for most purposes.

\section{DDS update rate and parametric heating}
\label{sec:parametric_heating}

In addition to the comparison of velocity profiles, our experimental and theoretical studies can also be used to determine the effect of the DDS update rate $f_{\rm DDS}$ on parametric heating. From Fig. \ref{fig:results_main} we see that in order to avoid retention loss from parametric heating, the update frequency should be chosen to avoid the fundamental trap resonance and any of its subharmonics. Choosing $f_{\rm DDS} \gtrsim 1.5 f_{\rm trap}$ seems generally to be quite safe. Note that parametric resonance losses do not occur for $f_{\rm DDS} \approx 2 f_{\rm trap}$, as would be expected for heating caused by oscillations of the trap depth \cite{Jauregui2001}. This results from the fact that the velocity boosts are directional, so that discretization effects from multiple boosts performed over the course of one oscillation cycle tend to cancel. Also parametric interactions with the transverse motional frequencies are expected to be negligible, since the trap velocity changes only in the longitudinal direction.

\begin{figure}[t]
\includegraphics[width=3.4in]{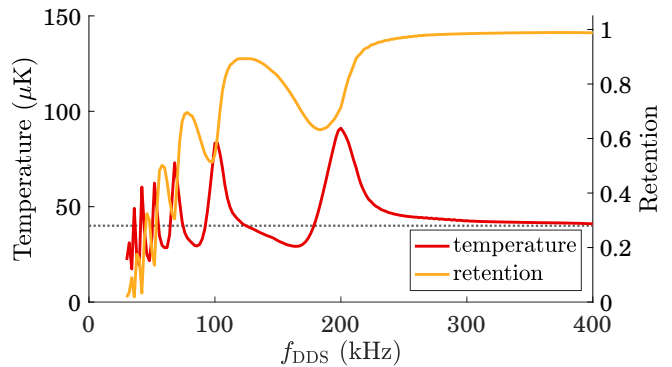}
\caption{Retention and final motional temperature for the simulation of Fig. \ref{fig:results_main}(a). Temperature increase due to parametric heating occurs only on the high frequency side of a parametric resonance. On the low frequency side we observe cooling instead. The horizontal dashed line indicates the initial motional temperature of the atoms.}
\label{fig:retention_and_temp}
\end{figure}

For the case in which one is concerned primarily with heating rates rather than retention losses, the situation can be slightly different. Figure~\ref{fig:retention_and_temp} illustrates the difference by plotting the simulated final motional excitation along with the simulated retention data from Fig. \ref{fig:results_main}. In order to produce the plot, the average motional energies are calculated from the final density matrices and then converted into an effective temperature.  The motional temperature increases only on the high frequency side of each parametric resonance, while cooling occurs on the low frequency side. This results from the anharmonicity of the trap and was used in \cite{Poli2002} to cool a cloud of trapped atoms by forced parametric excitation. If the goal is to to avoid an increase of motional temperature, either the DDS update frequency should be chosen such that $f_{\rm DDS} \gtrsim 1.5 f_{\rm trap}$, or it may be placed near the low-frequency side of a parametric resonance in retention loss.

Retention is an important figure of merit for transport systems in general. However in more recent experiments it has become increasingly important to maintain a high population fraction of atoms in the motional ground state of a trap \cite{Hamann1998, Meng2018, Wang2019, Levine2019, Graham2019}. In some experiments it will be possible to perform ground-state cooling after transport has been completed, so that maintaining a high ground state population during transport will be unnecessary. However this will not always be the case. For instance it may be that optical access to the site of a particular experiment is limited, or that other atoms or quantum systems are present at the experiment site that would be disturbed by the cooling process. Future work for our experiment includes plans to implement ground-state cooling, though at present our system does not have the ability to probe motional ground state populations. In the meantime these studies can be done using our theoretical model.

\section{Ground state populations}
\label{sec:state_population}

\begin{figure}[t]
\includegraphics[width=3.4in]{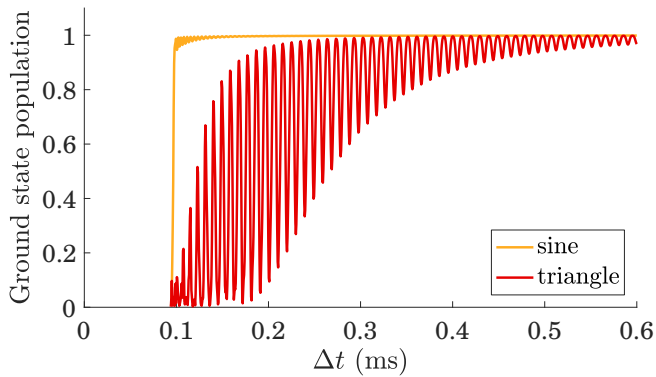}
\caption{Simulated final ground state populations after an atom is initialized in the motional ground state and then moved over a distance of $\Delta x=0.2$ mm. Simulation parameters are the same as those used for Fig.~\ref{fig:results_main}(b).}
\label{fig:state_population}
\end{figure}

Having validated our model empirically, we now use it to study ground state populations of atoms transported in an optical conveyor. The situation of interest to us is that of an atom which is cooled to the motional ground state of an optical lattice in one position, and then transported to another position for use in an experiment. For convenience we take the transport distance to be equal to the one used in the previous section, $\Delta x=0.2$ mm. We repeat the simulations shown in Fig.~\ref{fig:results_main}(b), but rather than using a thermal state as the initial condition we assume that the atom begins in the longitudinal motional ground state. Also, this time the atom is moved in a single direction only once, rather than being shuttled back-and-forth ten times, since this is the situation relevant for experiment. For a performance metric we use the final ground state population after transport. The results of simulations using triangle and sine velocity profiles are shown in Fig.~\ref{fig:state_population}.

Ground state population is exceptionally well preserved by the sine velocity profile, until the maximum acceleration $a_{\rm max}$ is reached with $\Delta t =0.094~$ms. By comparison the triangle profile performs quite poorly. Discontinuities in the acceleration perturb the state at three separate points in the transport procedure. Interference between these perturbations introduces population oscillations between the ground and first few excited states. These oscillations can lead to an increased sensitivity to slight changes in optical trap parameters, and even at an interference maximum the triangle waveform is still outperformed by the sine.

The oscillations apparent in Fig. \ref{fig:state_population} do not appear in Fig. \ref{fig:results_main}(b) for two reasons. Firstly, the anharmonicity of the trap tends to prevent any coherent population oscillations from being directly observed in retention data. Secondly, the dephasing rates of the higher-energy motional states in our model are much larger than those of the less highly excited states. Dephasing of the higher-lying states tends further to suppress the visibility of any population oscillations in the retention.

\begin{figure}[t]
\vspace{0.5cm}
\includegraphics[width=3.4in]{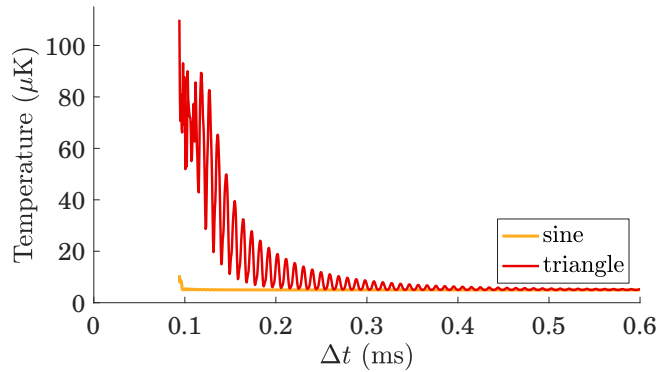}
\caption{Simulated final motional temperature after an atom is initialized in the motional ground state and then moved by $\Delta x=0.2$ mm. Parameters are the same as those of Fig. \ref{fig:results_main}(b). The energy between adjacent trap eigenstates corresponds to about $10~\mu$K in this example.}
\label{fig:motional_temp}
\end{figure}

Using data from the simulation of Fig. \ref{fig:state_population}, we can also compare the amount of atomic motional excitation acquired by the two velocity profiles. The results are shown in Fig. \ref{fig:motional_temp}. The sine shape causes very minimal excitation, whereas the triangle excites the atom by up to $100~\mu\rm K$, roughly equivalent to 10 times the energy between adjacent trap states. For longer transport durations though the heating caused by either profile becomes negligible. So if experimental circumstances allow motion to be accomplished slowly enough, then a triangle-shaped waveform may still be adequate.

\section{Conclusion}
\label{sec:conclusion}

In conclusion, we have performed experimental and theoretical studies of retention and heating in an optical conveyor belt. We have compared the performances of triangle-shaped and smooth velocity profiles, and also briefly examined parametric losses and heating caused by the DDS update rate. We found that a smooth, sine-shaped profile performs significantly better than does a triangle shape. The sine profile noticeably increases retention and decreases motional excitation for short transport times, approaching the absolute limit imposed by the depth and geometry of the trap. Using our theoretical model we showed that the performance advantage of the smooth sine profile is even greater when transporting atoms in the motional ground state. A range of other smooth velocity profiles was studied as well, besides the sine shape, but these were not included in this work since their performance was found to be very similar to that of the sine. The degree of motional excitation seems to have only weak dependence on the details of the velocity waveform used, as long as the changes in acceleration occur smoothly from the perspective of the atom.

This research was supported by the US Army Research Laboratory Center for Distributed Quantum Information through Cooperative Agreement No. W911NF-15-2-0061. We thank ColdQuanta, Inc. for loan of the 1064 nm laser. We also thank Matthew Ebert for assistance installing ARTIQ, and Timothy Ballance for help with DDS RAM mode.

\end{document}